\documentclass[english]{article}
\usepackage[T1]{fontenc}
\usepackage[latin9]{inputenc}
\usepackage{textcomp}
\usepackage{amsmath}
\usepackage{amssymb}
\usepackage{graphicx}

\makeatletter
\usepackage{babel}
\usepackage{cite}
\usepackage{hyperref}
\hypersetup{colorlinks=true,citecolor=blue}

\makeatother

\usepackage{babel}
\begin{document}

\title{Axigluon Phenomenology using ATLAS dijet data}

\author{Bastián Díaz and Alfonso R. Zerwekh%
\thanks{alfonso.zerwekh@usm.cl%
}\\
 \textit{\normalsize Departamento de Física and }{\normalsize }\\
{\normalsize{} }\textit{\normalsize Centro Científico-Tecnológico de
Valparaíso}{\normalsize }\\
{\normalsize{} }\textit{\normalsize Universidad Técnica Federico Santa
María}{\normalsize }\\
{\normalsize{} }\textit{\normalsize Casilla 110-V, Valparaíso, Chile}}
\maketitle
\begin{abstract}
In recent years, there has been a renewed interest on axigluons as
part of a possible extension of strong interactions at high energies.
In this work, we use recent ATLAS measurements of the dijet spectrum
in order to set limits on the axigluon mass and coupling to quarks.
We pay special attention to the methodology used to extract the resonant
contribution from theoretical simulations. Finally, we present some
predictions for the next LHC run at $\sqrt{s}=14$ TeV. 
\end{abstract}

\section{Introduction}

The Standard Model (SM) has proven to be extremely successful describing
collider data. Even the discovery of the new 125 GeV scalar state\cite{Aad:2012tfa,Chatrchyan:2012ufa},
which seems to behave like the so-long awaited Higgs boson, has unexpectedly
confirmed the validity of the SM at the Fermi scale. At the moment,
we don't have any collider hint of New Physics except, maybe, the
anomaly on the forward-backward asymmetry in the top--anti-top production
($A_{FB}^{t\bar{t}}$) observed at the Tevatron\cite{Aaltonen:2008hc,Aaltonen:2011kc,Abazov:2007ab}.
One of the possible explanations of this phenomenon is the existence
of a color-octet spin-1 field with axial coupling to quarks, the so
called axigluon. Many particular axigluon models have been advanced
in order to explain the $A_{FB}^{t\bar{t}}$ anomaly\cite{Ferrario:2008wm,Martynov:2009en,Ferrario:2009bz,Frampton:2009rk,Cao:2010zb,Bai:2011ed,Zerwekh:2011wf,Gresham:2011pa,Djouadi:2011aj,Haisch:2011up,Tavares:2011zg,AguilarSaavedra:2011ci,Krnjaic:2011ub,Wang:2011hc,Cvetic:2012kv}.
On the other hand, at the LHC, ATLAS\cite{Aad:2011aj,Aad:2011fq}
and CMS\cite{Khachatryan:2010jd,Chatrchyan:2011ns,CMS:2012yf,Chatrchyan:2013qha}
have performed axigluon searches in the dijet spectrum although considering
only the minimal model. Moreover, there has been some controversy
about the correct procedure for comparing the experimental data and
theoretical computations\cite{Harris:2011bh}. In this work, after
recalling the theoretical framework for axigluon physics (section
\ref{sec:Theoretical-Frammework}) we use recent ATLAS dijet data
and update the limits on the axigluon mass for the minimal universal
model (section \ref{sec:Limits-on-Axigluon-Mass}). More importantly,
in \ref{sec:Limits-on-Axigluon-Couplings} we reinterpret the existing
data in order to set limits on the axigluon coupling to quarks. In
this process, we pay special attention to the methodology used to
extract the resonant contribution from theoretical simulations. In
section \ref{sec:Axigluon-at-14TeV} we present some predictions for
the next LHC run at $\sqrt{s}=14$ TeV. Finally, in section \ref{sec:Summary-and-Conclusions}
we summarize and state our conclusions.

\section{Theoretical Framework\label{sec:Theoretical-Frammework}}

The axigluon arises in models where the strong interaction gauge group
is extended in order to provide a non-trivial chiral structure to
it\cite{Frampton:1987ut,Frampton:1987dn}. This kind of models are
generally called chiral color models. The minimal example is to consider
the group $SU(3)_{L}\times SU(3)_{R}$ and assume that left-handed
quarks transform under fundamental representation of $SU(3)_{L}$
while right-handed quarks are triplets of $SU(3)_{R}$. This symmetry
is then broken to the diagonal group which is identified with the
usual color group. After the symmetry breaking, it is found in the
physical spectrum a massless gauge boson with vector coupling to quarks
(the gluon) and a massive spin-1 field with axial-vector coupling
to quarks, the so called axigluon (which we will denote by $A_{\mu}$).
If we assume that the coupling constants associated with the left
group ($g_{L}$) and the right group ($g_{R}$) happen to be equal
($g_{L}=g_{R}=g$) then the usual QCD coupling constant (defined by
the coupling of quarks to gluon) is given by $g_{QCD}\equiv g/\sqrt{2}$.
In this simple scenario, which we call the Minimal Axigluon Model,
the axigluon has a pure axial coupling and its associated coupling
constant is just $g_{QCD}$. However, if $g_{L}\neq g_{R}$ the coupling
of the axigluon to quarks is described by the following Lagrangian\cite{Martynov:2009en}:

\begin{equation}
\mathcal{L}_{A\bar{q}q}=g_{QCD}\bar{q}\frac{\lambda^{a}}{2}\gamma^{\mu}\left(\cot(\theta)+\frac{1}{\sin(2\theta)}\gamma_{5}\right)A_{\mu}^{a}q\label{eq:Aqq_nominimo}
\end{equation}
where $\theta$ is a mixing angle defined by $\tan(\theta)=g_{R}/g_{L}$.
Notice that in this case the axigluon interaction has a vector part
as well as an axial-vector part and the coupling constant of the axial-vector
part is greater than $g_{QCD}$. Additionally, it is easy to see that
in this scenario the axigluon interaction is universal.

It is possible, however, to extend these ideas in order to obtain
a more flexible axigluon with modified and even non-universal couplings.
Two mechanisms are described in the literature that make possible
such an extension. The first one consists in introducing new heavy
exotic vector-like quarks to which normal quarks can mix up\cite{Bai:2011ed}.
The axigluon coupling to (normal) quarks is then modified by the presence
of functions of a new mixing angle in the quark sector. This mixing
may be flavor dependent and, hence, universality is broken. The second
mechanism is based on the introduction of extra spin-1 color octet
fields\cite{Zerwekh:2009vi}. A simple realization of this idea is
to extend the chiral color group to $SU(3)_{1}\times SU(3)_{2}\times SU(3)_{3}\times SU(3)_{4}$
and, using the ideas of Deconstruction Theory, ``delocalize'' quarks
in the different groups\cite{Zerwekh:2011wf}. After the symmetry
breaking process, and using an appropriated delocalization pattern,
the axigluon interaction is modified by the delocalization parameter.
Of course, each flavor of quark may be delocalized in a different
way and, as a consequence, universality is lost again.

In both mechanisms, the resulting axigluon coupling constant can be
arbitrary large or small depending on the value of the mixing or delocalization
parameter even in the case where the coupling constants associated
to the different groups are all equal.

In this work, we adopt a phenomenological approach and we consider
an axigluon interaction term of the form:

\begin{equation}
\mathcal{L}_{A\bar{q}q}=kg_{QCD}\bar{q}\frac{\lambda^{a}}{2}\gamma^{\mu}\gamma_{5}A_{\mu}^{a}q\label{eq:Aqq-Pheno}
\end{equation}
where $k$ is a (flavor independent) constant which measure the deviation
from the minimal model.

\section{Limits on Axigluon Mass\label{sec:Limits-on-Axigluon-Mass}}

\subsection{Methodology}

In recent years, ATLAS and CMS have searched for new resonances in
the dijet spectrum. One of them is the axigluon%
\footnote{Unfortunately, it seems that ATLAS has not updated its limits on axigluons
since 2011%
}. Unfortunately, a proton-proton collider like the LHC is not the
best place to search for axigluons. The reason is that an axigluon
resonance can only be produced by quark--anti-quark annihilation,
but this mechanism is suppressed at the LHC due to a parton distribution
function (PDF) conspiracy. That means that in a proton-proton collider
it is difficult to find anti-quarks with large momentum and therefore
it is difficult to produce a very massive axigluon on-shell. The consequence
of this suppression is that a large amount of dijet events originated
by an axigluon in the s-channel are in fact produced at low dijet
invariant masses by an off-shell axigluon. Nevertheless, in actual
measurements, this large tail is hidden under the usual QCD background.
For the theoretical side, it means that only the resonant part of
the cross section must be extracted from any simulation destined to
be compared to experimental data. ATLAS and CMS have used different
criteria for such an extraction in their setting up of limits for
the axigluon mass\cite{Harris:2011bh}. CMS made their simulation
relaying on the well known Narrow Width Approximation. ATLAS, on the
other hand, computed the leading order cross section considering all
the possible channels (not only the s-channel) and then integrated
the differencial cross section in the neighborhood of the resonance
(specifically in the interval $\left[0.7M,1.3M\right]$ where $M$
is the axigluon mass)\cite{Aad:2011aj}. These different criteria
lead to significantly different limits which originate some level
of controversy.

In this work, however, we implement a different procedure. Using CalcHEP\cite{Belyaev:2012qa},
we generate events for dijet production due to axigluon exchange taking
into account s-channel and t-channel contributions. Then, after applying
the appropriated kinematic cuts, we fit the dijet invariant mass cross
section distribution around the peak by using a Breit-Wigner plus
a second order polynomial. We define the extracted resonant cross
section as the integral of the Breit-Wigner function. Of course, this
procedure is equivalent to the Narrow-Width Approximation for narrow
resonances produced in the s-channel but is generalizable to the case
of wider resonances.

\subsection{Results}

\begin{figure}
\begin{centering}
\includegraphics[scale=0.5]{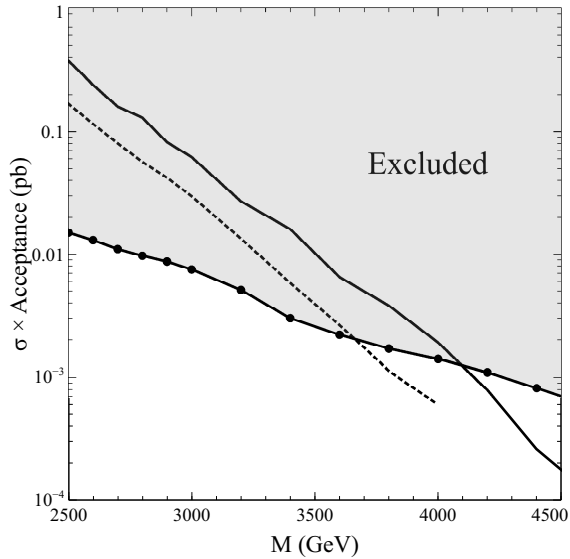} 
\par\end{centering}

\caption{Predicted resonant cross sections with (continuous line) and without
(dashed line) detector resolution compared to ATLAS upper limits to
narrow resonances decaying into dijet (dots with continuous line).
In both cases we used $k=1$ \label{fig:Mass-Limits}}
\end{figure}

Usually, experimental collaborations set limits on axigluon mass based
only on the Minimal Axigluon Model (that is, $k=1$ in (\ref{eq:Aqq-Pheno})).
In order to compare our results with existing limits, as a first step,
we focus in this simple case. As explained above, we use CalcHEP to
generate events for dijet production due to axigluon exchange. In
order to compare to ATLAS measurements of the dijet spectrum, we impose
the following cuts:

\begin{eqnarray*}
|y_{i}| & < & 2.8\quad(i=1,2)\\
|y_{1}-y_{2}| & < & 1.2\\
M_{jj} & > & 1000\:\mathrm{GeV}\\
p_{Ti} & > & 150\:\mathrm{GeV}\;(i=1,2)
\end{eqnarray*}
where $y_{i}$ and $p_{Ti}$ are the rapidity and transverse momentum
of each jet while $M_{jj}$ represents the invariant mass of the dijet
system. After extracting the resonant part of the cross section, as
explained above, we compare our results with the limits obtained by
ATLAS for resonances decaying into dijets at $\sqrt{s}=8$ TeV and
$\mathcal{L}=$13 fb$\!^{-1}$\cite{ATLAS:2012qjz}. The result is
shown in figure \ref{fig:Mass-Limits}. The continuous line represents
our extracted resonant cross section for different values of the axigluon
mass. The line with dots is the experimental limit set by ATLAS for
resonances. Every resonance producing a cross section above the experimental
limit is excluded. We see that, in the context of the Minimal Axigluon
Model, that the axigluon is excluded if it is lighter that $4100$
GeV. This limit is comparable with recent limits obtained by CMS\cite{CMS:kxa}.

Our second step is to investigate the influence of detector effects
on our limit. In our case, we focus on the resolution of the calorimeter.
A typical resolution for a hadron calorimeter is:

\begin{equation}
\frac{\Delta E}{E}=\mathrm{\frac{0.5}{\sqrt{E(GeV)}}}\oplus0.03
\end{equation}

We smeared our events using this parametrization for the resolution
of the calorimeter and we re-extracted the resonant cross section
using the same cuts. Of course, due to the smearing procedure we fitted
the peak with a gaussian function and not a Breit-Wigner one. Our
new predicted cross sections are represented in figure \ref{fig:Mass-Limits}
by the dotted line. We see that this time axigluons are excluded if
they are lighter than $3700$ GeV.

\section{Limits on Axigluon Couplings\label{sec:Limits-on-Axigluon-Couplings}}

\subsection{Narrow Axigluon}

A limitation of the usual axigluon searches is that it is based on
the Minimal Axigluon Model. That means that it is always assumed that
the axigluon couples to quarks with QCD intensity. Nevertheless, this
is far from being the general case. In fact, it is a simple but very
special scenario. For this reason, we reinterpret the ATLAS limits
in order to constrain the coupling of the axigluon to quarks.

We assume that the axigluon interaction with quarks is given by Lagrangian
(\ref{eq:Aqq-Pheno}). If the axigluon is relatively narrow the following
relation among cross sections must hold:

\begin{equation}
\sigma_{k\neq1}=k^{2}\sigma_{k=1}
\end{equation}

On the other hand, if the axigluon has remained invisible it must
happen that:

\begin{equation}
\sigma_{k\neq1}\leq\sigma_{\mathrm{ATLAS}}
\end{equation}
where $\sigma_{\mathrm{ATLAS}}$ represents the limit on cross section
imposed by ATLAS searches. Thus, we can extract limits on the parameter
$k$:

\begin{equation}
k^{2}\leq\frac{\sigma_{\mathrm{ATLAS}}}{\sigma_{k=1}}
\end{equation}

\begin{figure}
\begin{centering}
\includegraphics[scale=0.5]{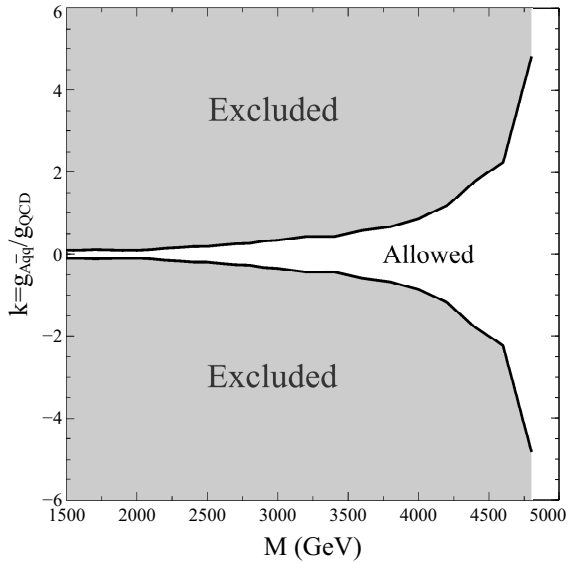} 
\par\end{centering}

\caption{Limits on the $k$ factor defined as $|k|=g_{A\bar{{q}q}}/g_{\mathrm{QCD}}$\label{fig:k-limits}}
\end{figure}

The resulted limits are shown in figure \ref{fig:k-limits}.

Axigluons with masses ranging in the $\left[1.5,2.5\right]$ TeV interval
are still allowed provided that they are fairly weakly coupled, with
$|k|$ in the $\left[0.1,0.2\right]$ interval.

\subsection{Broad Axigluon}

\begin{figure}
\begin{centering}
\includegraphics[scale=0.5]{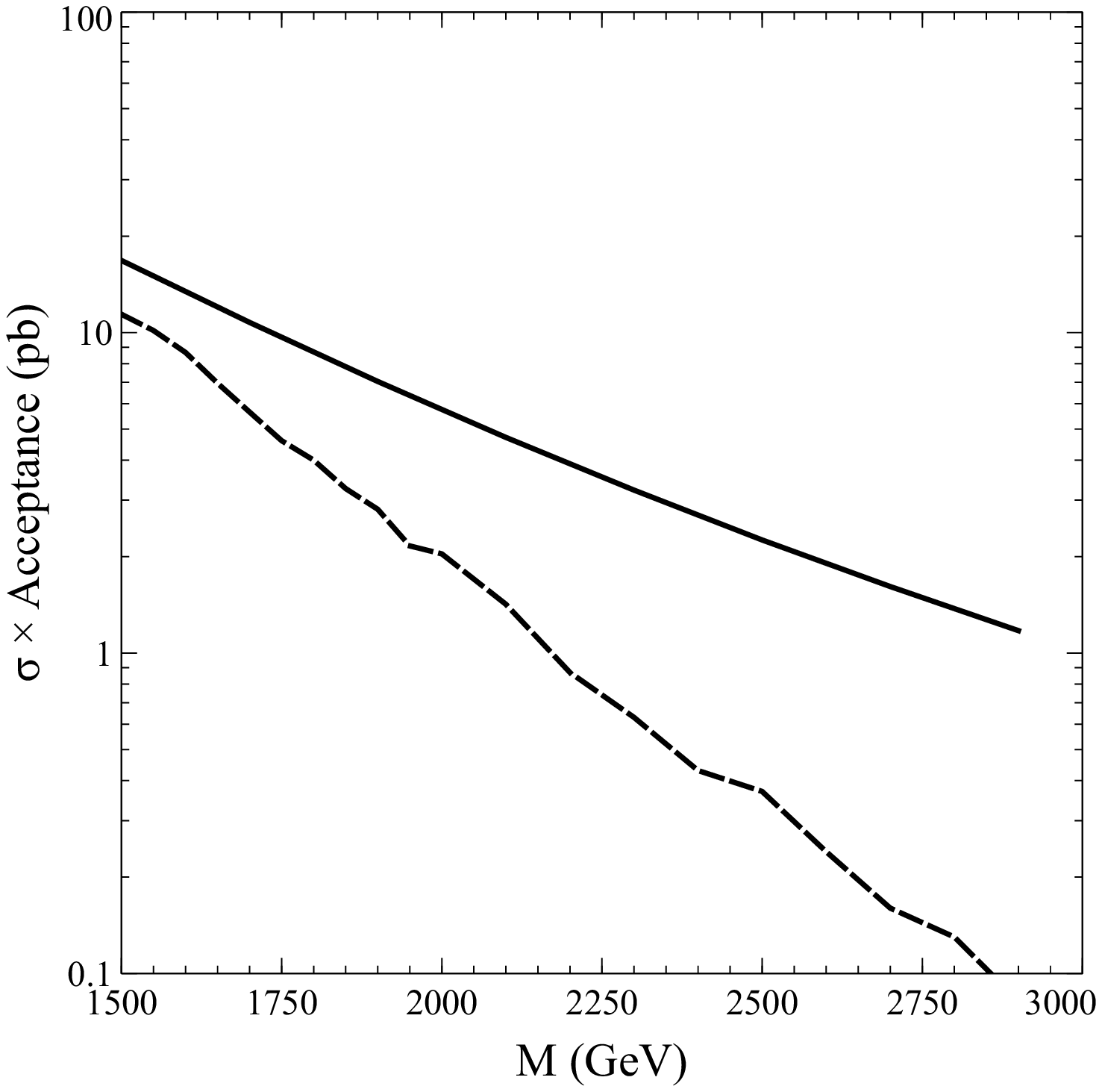} 
\par\end{centering}

\caption{Resonant cross section for a broad axigluon ($\Gamma/M=0.3$, continuous
line) compared to the resonant cross section for an axigluon decaying
into standard quarks. In both cases we used $k=1$. \label{fig:Xsection-Large}}
\end{figure}

So far, we have considered an axigluon which decays only into standard
quarks with universal coupling. This kind of axigluon is relatively
narrow with $\Gamma/M\sim0.1$. Nevertheless, phenomenologically it
is more important the case of a broad axigluon ($\Gamma/M\gtrsim0.2$)
since it is a viable explanation for the $A_{FB}^{t\bar{t}}$ anomaly
observed at the Tevatron. An axigluon may be broad if its width is
dominated by a strong decay channel into non-standard particles or
even to the top-quark. We tried to remain model-agnostic and we studied
a broad axigluon in an illustrative case. Again, we generated events
for dijet production due to axigluon exchange but this time we set
the ratio $\Gamma/M$ to a fixed value: $\Gamma/M=0.3$. For axigluon
heavier than $3$ TeV it is impossible to fit any resonant structure
and consequently the axigluon is unobservable by searches in the dijet
spectrum. However, for masses in the range $1$ TeV $\leq M<3$ TeV,
resonant structures are recognizable and the resonant cross sections
can be extracted. In figure \ref{fig:Xsection-Large}, we compare
the extracted resonant cross section for the broad axigluon (continuous
line) with the one obtained previously for a narrow axigluon (dashed
lined). In both cases we used $k=1$. We see that the broad axigluon
produces larger cross sections. This is mainly due to the fact that,
in the broad axigluon case, the axigluon resonance explores lower
invariant masses suffering less PDF suppression and thus enhancing
the integrated cross section.

\section{Axigluon at $\sqrt{s}=14$ TeV\label{sec:Axigluon-at-14TeV}}

\begin{figure}
\begin{centering}
\includegraphics[scale=0.5]{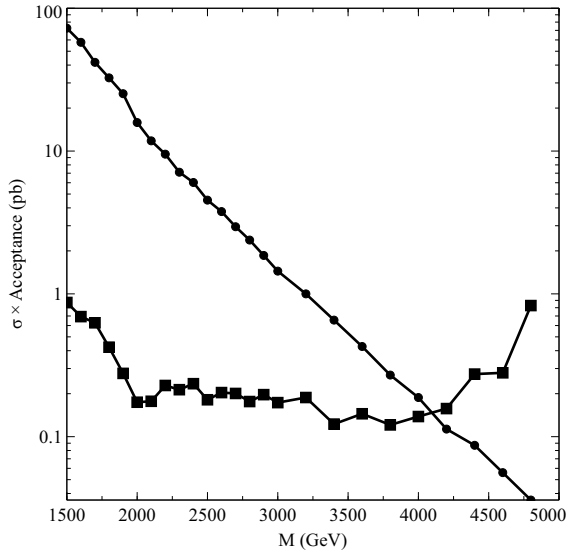} 
\par\end{centering}

\caption{Resonant cross section predicted for the 14 TeV LHC using $k=1$ (dots
with continuous line) and the maximum value of $k$ allowed by current
data (squares with continuous line) \label{fig:Prediction}}
\end{figure}

We now move toward our expectations for the next run of the LHC at
$\sqrt{s}=14$ TeV. In figure \ref{fig:Prediction} two curves are
shown. The first one (dots with continuous line) represents the resonant
cross section predicted by the Minimal Axigluon Model. The second
one, (squares with continuous line) represents the resonant cross
section obtained considering the maximum value of $k$ allowed by
current data for each axigluon mass. Our results tell us that if a
(narrow) axigluon exists but has escaped our searches because it is
couple to quarks weakly enough, then it would be produced at the 14
TeV LHC with maximum resonant cross section laying somewhere in the
range $0.1-1$ pb.

\section{Summary and Conclusions\label{sec:Summary-and-Conclusions}}

The axigluon, although it may be seen as an exotic kind of New Physics,
has an interesting phenomenology and may be a natural explanation
of an observed, and still not understood, anomaly. In this work, we
have used recent ATLAS measurements of the dijet spectrum to set limits
on the mass and couplings of the axigluon in a more general theoretical
context than the one usually consider by experimental searches. The
coupling of axigluon with mass below $3$ TeV to light quarks is highly
constrained for narrow and broad axigluon. In contrast, a broad axigluon
with $\Gamma/M=0.3$ is eventually invisible in the dijet spectrum
for masses larger than $3$ TeV . Additionally, we have defined a
methodology for extracting the resonant cross section which is generalizable
beyond the narrow resonance case.

\section*{Acknowledgement}

ARZ has received financial support from Fondecyt grant nº 1120346

\end{document}